\documentclass{osa-article}
\journal{boe}
\usepackage{lineno}
\begin{document}

\title{Statistical Estimation of Ballistic Signal in Visible Light OCT Based on Random Matrix Description}

\author{DANLEI QIAO,\authormark{1,2,4} PENG MIAO,\authormark{1,4,*} IAN RUBINOFF, \authormark{2}JIBO ZHOU,\authormark{3} JOHN B. TROY,\authormark{2} HAO F. ZHANG,\authormark{2} AND SHANBAO TONG\authormark{1,\dag}}

\address{\authormark{1}School of Biomedical Engineering, Shanghai Jiao Tong University, Shanghai 200240, China\\
\authormark{2}Department of Biomedical Engineering, Northwestern University, Evanston, IL, USA\\
\authormark{3}Department of Ophthalmology, Shanghai Ninth People's Hospital, Shanghai Jiao Tong University School of Medicine, Shanghai 200011, China\\
\authormark{4}These authors contribute equally to this work}
\email{\authormark{*}pengmiao@sjtu.edu.cn;\authormark{\dag}stong@sjtu.edu.cn} %% email address is required

% \homepage{http:...} %% author's URL, if desired

%%%%%%%%%%%%%%%%%%% abstract %%%%%%%%%%%%%%%%
%% [use \begin{abstract*}...\end{abstract*} if exempt from copyright]

\begin{abstract}
Visible light optical coherence tomography (vis-OCT) provides a unique tool for imaging the structure and oxygen metabolism in tissues. However, since it works in the spectral domain, vis-OCT still suffers from noises due to the multiple scatterings, \emph{e.g}. for imaging the human fundus. In this study, we modeled the OCT signals as a hybrid of single and multiple scattering components using Wishart random matrix description, with which the single scattering component thus can be separated out using the low-rank characteristics of the matrix. The model was validated using Monte Carlo simulation. We further demonstrated that this model could significantly improve the imaging performances in human fundus, showing more details of the vascular structure than the current vis-OCT and an increase of signal-to-noise ratio (SNR) up to more than 10dB. The layer structure of the retina can be better revealed with more than 3dB suppression of the blood scattering in OCT signals.
\end{abstract}

%%%%%%%%%%%%%%%%%%%%%%%%%%  body  %%%%%%%%%%%%%%%%%%%%%%%%%%
\section{Introduction}
Optical coherence tomography (OCT) has been widely applied to clinical diagnostics. It utilizes the coherence gating strategy to recover the 3D structural and blood flow information of layered tissue. In addition, visible light optical coherence tomography (vis-OCT) provides both anatomy and blood oxygen saturation ($sO_2$) information\cite{2013Visible,2017Retinal}. Another benefit of vis-OCT is the significant improvement of spatial resolution at the shorter wavelengths. These unique properties are critical for clinic diagnostics. For example, histopathology suggests that early ocular changes in retinal diseases including glaucoma, diabetic and age-related macular degeneration (AMD) occur at the micron scale of the multilayer structure between the retinal pigment epithelium (RPE) and Bruch’s membrane, that mediates RPE and choroidal capillaries metabolism transport.

Ballistic signal, single scattering, is desirable in OCT imaging, but the disturbances due to multiply scattering from remote and deeper tissue reduces its sensitivity and accuracy\cite{1991Optical,2020Deep}. The interference by the multiply scattered light is more profound in spectral domain OCT (SD-OCT), e.g. vis-OCT, owing to the low coherence of the light source. Compared with traditional OCT, vis-OCT works at shorter wavelength and provides a higher spatial resolution. Nonetheless, visible light would experience more scattering in tissues due to its smaller scattering mean free path \emph{$l_s$} than that of near-infrared light, which thus degrades the signal-noise-ratio (SNR) of the OCT signals.

New strategies for separating singly and multiply scattered light have been proposed, such as wave-front shaping with a spatial light modulator (SLM) and matrix measurements for light propagation through complex media\cite{2016Smart}. Nevertheless, SLM based method has complex procedures and slow imaging speed, which limits its clinical or other \emph{in vivo} applications. Mathematical model thus can be an alternatively way to statistically describe single scattered signal so as to efficiently separate the different scatterings in \emph{in-vivo} imaging setting.

In this study, we established the statistical description of single and multiple scattering components in vis-OCT signals according to random matrix theory (RMT)\cite{2009Random}. With this model, we further proposed a new estimation method for the single-scattering components (ballistic photons). Monte Carlo simulation was implemented to validate the model and test the improvements of signal-noise ratio (SNR) and contract-noise ratio (CNR). In an application to human retina vis-OCT imaging, we compared the improvement of SNR and CNR after implementation of RMT to the traditional vis-OCT, as well as the state of art post-processing methods.

\section{Theory}
\subsection{Signal formation in vis-OCT}
The standard signal formation modal for vis-OCT system is based on the field-field interference of the reference arm ($E_R(k)=Re\{\sqrt{I_0(k)}e^{-i(kL_0-\omega t)}\}$) and the sample arm ($E_S(k)=Re\{u(k)\sqrt{I_0(k)}e^{-i(\Phi_S(k)+kL_0-\omega t)}\}$). $I_0(k)$ is the source power spectrum, $k = 2\pi/\lambda$ is wave number, $L_0$ is the double arm length, $\Phi_S(k)$ is the additional phase argument due to the scattering inside the sample. For the ideal scenarios, \emph{i.e.} purely ballistic trajectories with single scattering, $\Phi_S(k)=k\Delta L$ with $\Delta L=2z$ and $z$ is the depth position of the back-scattering event. The stationary spectrum of the vis-OCT is: 
\begin{equation}
I(k) = [E_R(k)+E_S(k)]^2 = I_0(k)[1 + u^2(k) + 2u(k)Re\{e^{-ik\Delta L}\}].  \\
\label{con:I_K} 
\end{equation}

We then removed its \emph{dc} part and obtained the pre-processed spectrum $\tilde{I}(k)$ after a  re-normalization, which was then used to recover the depth-resolved scattering amplitude $S(\Delta L)$ by inverse Fourier transform. In practice, we use the line-CCD in spectrometer to record the signals $I(\lambda_n)$ in which $n$ is the pixel number at the specific wavelength $\lambda_n$. Considering the relation of $k = 2\pi/\lambda $, we recorded $I(k_n)$ with $n = 1...N$. However, the sampling in the \emph{k}-space is not equally spaced which forbids the standard Fourier transform. An up-sampling procedure is thus performed to obtain the equally spaced $\tilde{I}(k_n)$ ($n = 1...T$) to reconstruct the $S(\Delta L)$ signal:
\begin{equation}
S(\Delta L)=\sum^{T-1}_{n=0}\tilde{I}(k_n)e^{2\pi i \Delta L(n/T)},
\end{equation}

For realistic imaging, both single and multiple scatterings contributes to $E_S(k)$. We can thus represent the $E_S(k)$ as the sum from all $M$ trajectories: $E_{j}(k)=u_j(k)e^{ikl_j}$ ($j=1,...,M$), in which $u_j(k)$ represents the field magnitude and $l_j$ is the scattering path length in the sample. $E_S(k)$ can thus be further separated into single and multiple scattering components: $E_S(k)$ = $E_{SS}(k)$ + $E_{SM}(k)$. With an assumption of $M_1$ single scattering trajectories, the first part can have the representation:
\begin{equation}
E_{SS}(k)= Re\{\sqrt{I_0(k)}\sum^{M_1}_{j=1}u_j(k)e^{ikl_j+kL_0-\omega t)}\}.
\end{equation}
The multiple scattering part is thus:
\begin{equation}
E_{SM}(k)= Re\{\sqrt{I_0(k)}\sum^M_{j=M-M_1+1}u_j(k)e^{ikl_j+kL_0-\omega t)}\}.
\end{equation}
The model for signal $I(k)$ is:
\begin{equation}
I(k) = I_{dc} + 2 Re\{I_0(k) \sum^{M}_{j=1} u_j(k)e^{ikl_j}\}.
\label{con:I_KNew} 
\end{equation}

Thus, the multiple scattering trajectories would distort the phase distribution,which becomes more profound in vis-OCT when all spectrum components are recorded simultaneously. Longer wavelength corresponds to smaller wave number. The multiple scattering of such wave packages satisfying $k_m L_{m}=k_l L_{l}$ forms the subgroup with same phase accumulation, which implies that multiple scattering disperses in the entire range of reconstructed signal $S(\Delta L)$. As a result, a specific coherence volume at location $\Delta L /2$, $S(\Delta L)$ contains both single and multiple scattering components and their interactions. The path length distribution determines the final $S(\Delta L)$, which would be sufficiently randomized and tend to be Gaussian with the increase of the scattering events.

The above procedure reconstructs one depth-resolved OCT signal, A-line. 3D imaging of tissue can be obtained by reconstruction of multiple B-scan, \emph{e.g.} using the $X-Y$ scanning galvanometer. The lateral scanning position is encoded as the data acquisition time, \emph{i.e.} $t\leftrightarrow (x,y)$. For a short time window $[t_0,t_1]$, the scanning positions are always closed to each other. 

\subsection{Wishart random matrix description}

For each A-line in vis-OCT of biological tissues, the structural inhomogeneity always exists in larger scale compared to the coherence length. This makes the single scattering components demonstrate low rank property. This low rank property is preserved in the neighbor A-lines when the inhomogeneity is spatially extended. For multiple scattering component, the low rank property disappears due to the Gaussian statistics. These specific conditions make the isolation of single scattering component possible. 

We use the output of spectrometer, \emph{i.e.} $I(n,t)$ with $n=1...N$ and $t\in[t_0, t_0 + (P - 1)\Delta t]$, to construct the $N \times P$ spectrum matrix $Z(n,p)=I(n,t_0+(p-1)\Delta t)$. Each column in $Z$ is the spectrum vector for an A-line. Similar to $I(k)$ (Eq.\ref{con:I_KNew}), $Z$ contains the single and the multiple scattering components: $Z=Z_S+Z_M$. The backscattered wavelets can be assumed statistically independent\cite{2003Optical}. Since $\lambda = 2\pi/k$, each element of $I(n)$ can be also considered as independent sampling in $k$ from the same spectrum distribution. When a sufficient short time window is applied to neighboring spatial window, different columns in $Z$ corresponds to samplings from the similar spectrum ensembles with the same structural inhomogeneity. Thus,the low rank characteristics for the single scattering part $Z_S$ are preserved in the $k$ space. Meanwhile, Gaussian distribution is also expected for the multiple scattering part $Z_M$, because under the Fourier transform of the Gaussian distribution dose not change itself distribution pattern. 

We further consider the eigenvalue distribution of Wishart RM$ H_M=\Tilde{Z}_M\Tilde{Z}_M^{'}$, where $\Tilde{Z}_M$ is the centralized matrix, \emph{i.e.} $\Tilde{Z}_M = Z_M - \Tilde{Z}_M^{'}$. $\Tilde{Z}_M^{'}$ is the averaged matrix along the row direction in $Z_M$. The corresponding hybrid matrix $H_H$ and single scattering matrix $H_S$ are defined in the same way. For a standard Wishart RM $H_M$, its eigenvalue distribution follows the Mar$\rm \check{c}$enko-Pastur law (MP Law) \cite{2009Random}. For the vis-OCT dataset, the spectrum dimension $N$ is always significantly larger than the sampling $P$, \emph{e.g.} $N=2048$ $\emph{v.s.}$ $P=7 $ in this study. The ratio of $N$ and $P$, \emph{i.e.} $\gamma = N/P$, determines the eigenvalue distribution of $H_M$ based on MP law. Under the conditions of $\gamma \gg 1$ and $N \rightarrow \infty$ , the MP law is a mixture of a point mass at $0$ and the density of standard MP distribution. Thus, for multiple scattering $H_M$, we have:
\begin{equation}
p_M(s) = (1-1/\gamma)\delta(s) + \frac{1}{2\pi\gamma\sigma^2_Ms}\sqrt{(b-s)(s-a)},\label{con:p_m},
\end{equation}
where $a = \sigma^2_M(1-\sqrt{\gamma})$ and $b = \sigma^2_M(1+\sqrt{\gamma})$ are the lower and upper boundary of eigenvalues, respectively.

The eigenvalue distribution of Wishart RM $H_H$, \emph{i.e.} $p_H (s)$, can be considered as the low rank single scattering component $p_S (s)$ biased by the multiple scattering component $p_M (s)$. The eigenvalue distribution of $H_S$ is bounded in a limited range due to the low rank property, \emph{i.e.} there are $Q$ significantly large eigenvalues with $Q \ll N$. Here we assume that the rank of $H_S$ is $Q$ in $k$ space. It should be noted that the ranks of single scattering in spatial space and $k$ space are usually different. Single scattering components also contribute to the larger eigenvalues in the $p_H (s)$. Usually, the single scattering contributes major energy in the signal of vis-OCT which means the eigenvalues near the upper boundary are good estimation for single scattering components. In the next section, we propose a separation method using proper criterions to isolate the eigenvalues majorly contributed by the low rank component. We further use Monte Carlo simulation to validate and evaluate the proposed method.

\subsection{Separation of eigenvalues for single scattering component }
The hybrid Wishart RM $H_H$ is symmetric and thus can be diagonalized with orthonormal matrix $U$ to obtain its eigenvalues {$s_n$}($n = 1\dots N$).
\begin{equation}
H_H = U{\varLambda}U',
\end{equation}
where $U$ is orthogonal matrix with size of $N\times N$. $\varLambda$ is a $N \times N$ diagonal matrix whose diagonal is the eigenvalues of $H_H$. We arrange the eigenvalues in descending order:$s_1\geq s_2\geq\dots\geq s_{N-1}\geq s_N$. 

Since RM demonstrates the eigenvalue repulsion, there is no exactly same eigenvalues. The smallest $N-P$ eigenvalues are asymptoticly close to zeros. The multiple scattering contribution exists in the full range of $\{s_1\sim s_P \}$, which, however,is relatively small for the largest eigenvalues (see the MC simulation result below), \emph{i.e.} $\{s_1,s_2,\dots,s_Q \}$, thus asymptotically closed to the single scattering components. 

The rank $Q$ of single scattering component can vary over A-lines since it is determined by the local structural inhomogeneity. We need an adaptive method to separate the $\{s_1,s_2,\dots,s_Q\}$. Here we propose a hypothesis testing method to reject all small eigenvalues in $\{s_{Q+1}\sim s_P \}$. This method uses the generalized likelihood ratio test (GLRT) to reject the small eigenvalues based on the statistic of $C$:
\begin{equation}
C = \frac{s_1}{\frac{1}{P}\sum^P_{n=2}s_n}\label{con:C},
\end{equation}
with the null hypothesis $H_0$: $C<s_{Q}$.
We note that both $P$ and $Q$ are smaller than $N$. Eq.\ref{con:C} works efficiently when $P\geq Q$. Furthermore, when the single scatterings component dominates, \emph{e.g.} the superficial layer, the largest $Q$ eigenvalues can practically estimate the single scattering components with good accuracy.The single scattering $Z_S$ matrix can be recovered by $\{s_1,s_2,\dots,s_Q\}$:
\begin{equation}
Z_S = U\sqrt{\hat{\varLambda}}\label{con:z_s}.
\end{equation}
where $\sqrt{\hat{\varLambda}}$ is a $N \times Q$ diagonal matrix with $\{\sqrt{s_1},\sqrt{s_2},\dots,\sqrt{s_Q}\}$ as its diagonal elements.

\section{Method}
\subsection{Monte Carlo simulation of vis-OCT }
We developed CUDA GPU accelerated Monte Carlo simulation platform base on Sherif’s work\cite{2017Massively}. One cross-section scan (B-scan) and 3D-scan of SD-OCT can be finished in 1 minute and 200 minutes, respectively. We make a 4-layer 3D phantom model of the posterior ocular structure in Monte Carlo simulation, as shown in Fig.\ref{fig:figmonte}(a). The 4 layers correspond to retina, retina pigment epithelium (RPE), choroid, and sclera with the respective thickness of $200, 10, 250, $and $ 700 \mu m$ referring to \cite{2013Accuracy}, with a horizontal cylindrical blood vessel segment deliberately placed in the retina layer. The vessel has a diameter of $130\mu m$ and wall thickness of $10\%$ of the lumen diameter.\cite{2009Wall}.  The blood within the vessel assumed to be optically homogeneous. The optical properties of solid tissues, including absorption coefficient $\mu_a$ [$cm^{-1}$], scattering coefficient $\mu_s$ [$cm^{-1}$], and anisotropy factor $g$ [dimensionless], are referring to \cite{Marleen1989Fluorescence, 1995Optical}, which are also assumed to be constants within the visible-light spectrum band (see Table \ref{con:phantom}).
\begin{table}
\centering
\caption{Optical parameters of the object}
\begin{tabular}{p{3cm}p{2cm}p{2cm}p{2cm}p{2cm}}
\arrayrulecolor{black}
\hline
Medium      & $\mu_a$  & $\mu_s$  & $g$     & $n$    \\
\hline 
Retina      & 1.47      & 31        & 0.97  & 1.37 \\
Vessel wall & 6.3       & 277       & 0.86  & 1.37 \\
Vessel      & 110.23    & 693.17    & 0.972 & 1.37 \\
RPE         & 938       & 1068      & 0.84  & 1.38 \\
Choroid     & 224       & 950       & 0.94  & 1.39 \\
Sclera      & 4         & 966       & 0.9   & 1.39\\ \hline
\end{tabular}\label{con:phantom}
\end{table}

We then launch a package of photons at the same initial location with a thin beam incident perpendicular to the surface of the tissue sample. The photons are traced according the rules in ref\cite{1995MCML,1999Monte} shown in Fig.1. The backscattered photons are collected by a single mode fiber with a collecting radius of 0.01 mm. Path lengths of the collected photons from depth $z$ can be sorted into two types of distributions : (1) ballistic lights from the photons that are scattered within 5 scattered events, and (2) multiply scattered photons.

\subsection{in-vivo OCT imaging of human retina}
\begin{figure}
\centering\includegraphics[width=12cm]{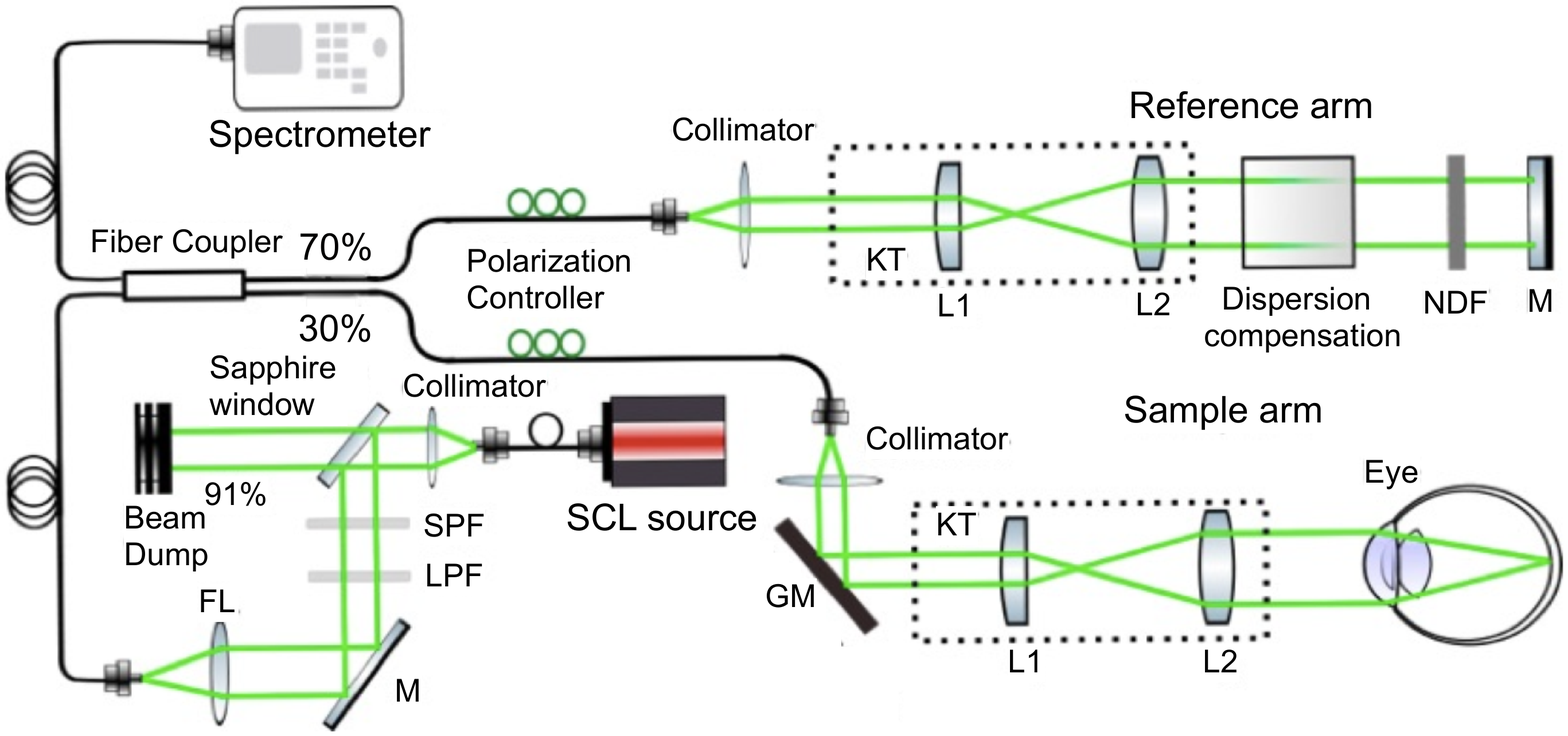}
\caption{Schematic of the vis-OCT system. GM, galvanometer mirrors; M, mirror; LPF, Longpass Filters; SPF,Shortpass Filters;L1,L2, lens; FL,Focal lens; SCL scource, supercontinuum light source, KT: Keplerian telescope (L1 and L2).}
\label{fig:OCTsystem}
\end{figure}

To demonstrate the W-RMT statistical seperation in real biomedical vis-OCT images, we apply our algorithms to the retina vis-OCT images collected with a prototype visible light OCT system developed at Northwestern University. The technical configurations of the system is shown in the Fig.\ref{fig:OCTsystem}. Briefly, A SD-OCT for in vivo imaging was built with a light from a supercontinuum light source (SuperK EXTREME; NKT Photonics, Birkerød, Denmark) that was then filtered and sent to a $30:70$ fiber coupler (Gould Fiber Optics, Millersville). A pair of galvanometer mirrors (Nutfield Technology, Londonderry) scanned the beam through an objective lens (LSM03, ThorLabs, Newton, which focused the light onto the sample. A visible light spectrometer (Blizzard SR; Opticent Health, Evanston, IL) operating from 510 to 610 $nm$ detected the interferogram signals for image reconstruction. The theoretical axial and lateral resolutions of the system in air were $1.3$ and $6.8$ $\mu m$, respectively. 

The experiment was approved by Northwestern University Institutional Review Board (IRB), and adhered to the tenets of the Declaration of Helsinki. All procedures took place in the Ophthalmology Department at the Northwestern Memorial Hospital. Healthy volunteers were recruited during their routine clinical visits. One or both eyes of each volunteer were imaged using both circular scanning and raster scanning. The former scans around the optic disc crossing all retinal vessels with 8192 A-lines, while the latter scans across the optic disc with 512 A-lines . Circular scanning thus takes 16 times of one raster scanning in time, which explains the low SNR in circular scanning OCT image.

\subsection{Metrics for evaluation}
In order to quantitatively evaluate the  single scattered signals by RMT analysis of OCT images, we implemented two image scores commonly used in image processing, \emph{i.e.}, the contrast-to-noise ratio(CNR) and peak SNR(PSNR).
CNR-\emph{dB}:
\begin{equation}
CNR = 20log_{10}\frac{\mu_s-\mu_b}{\sqrt{\sigma_s^2+\sigma_b^2}},
\end{equation}
where $\sigma_s$ and $\sigma_b$ are the standard deviation of the retina and background region, while $\mu_s$ and $\mu_b$ are the mean of the retina and background region.
PSNR (signal-to-noise ratio)-$dB$:
\begin{equation}
PSNR = 20log_{10}\frac{Max_{signal}}{\sigma_{b}},
\end{equation}
where $Max_{signal}$is maximum signal of the retina region.

\section{Results}
\subsection{MC simulation results}

\begin{figure}

\centering\includegraphics[width=12cm]{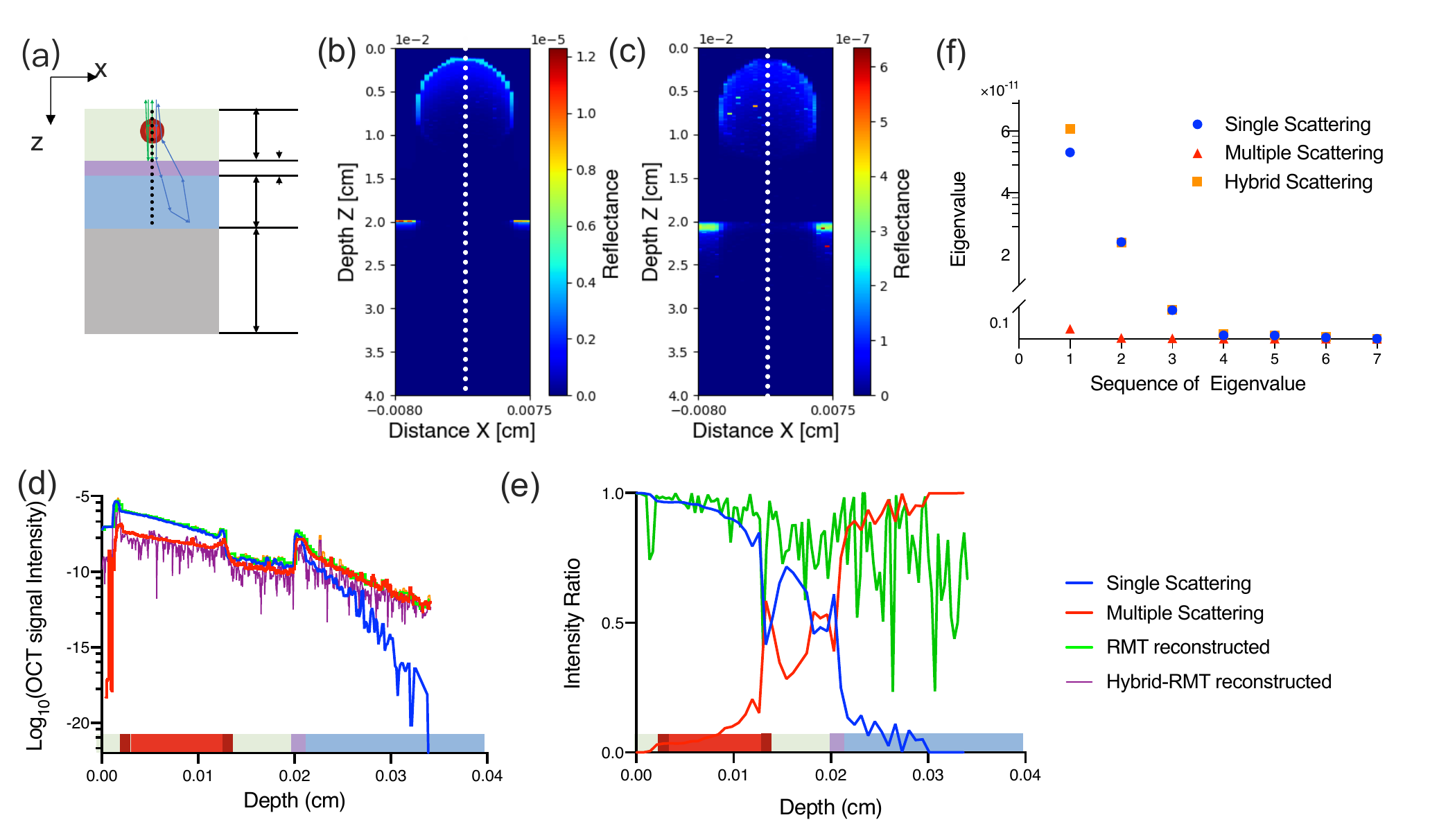}
\caption{Monte Carlo simulation. (a) fundus phantom in MC simulation.  (b) reconstructed OCT B-scan image only with ballistic photons. (c) reconstructed OCT B-scan image only with the multiple scattering photons. Dashed lines indicate the location x=0. (d) OCT A-lines along the lateral position x=0., (e) the ratio curves for OCT signals from single scatterings (blue) and multiple (red) scattering photons along the same A-line.  The color bars in (d,e) are corresponding to the layers defined in (a). , (f) Eigenvalues of the constructed Wishart RM using the OCT signals from ballistic, multiple scattering and hybrid scattering photons, respectively.}
\label{fig:figmonte}
\end{figure}

Fig.\ref{fig:figmonte}(a) shows a standard fundus phantom with 4-layer structures (retina, RPE , choroidal, sclera) and a blood vessel within the retina. The optical parameters in each layer are assumed to be distributed uniformly. Fig.\ref{fig:figmonte}(b-c) are the reconstructed B-scan images from the ballistic and multiple-scattering photons by MC stimulation, respectively. The ballistic B-scan image shows clear structure details of the vessel wall and layer boundaries. While B-scan image for the multiple scattering photons come out with much lower SNR. The distribution of scattering coefficients determines the strength of multiple scatterings. The high inhomogeneity of the blood vessel  results in more multiple scattering photons, which thus makes the vessel more visible in the  multiple scattering B-scan image(Fig.\ref{fig:figmonte}(c)). The overall ratio between ballistic (Fig.\ref{fig:figmonte}(b)) and multiple scattering (Fig.\ref{fig:figmonte}(c)) signals  is about $20:1$.  The MC simulation thus supports the idea that separation of the ballistic and multiple scattering signals would help to improve the $SNR$ compared with the original signals. 
Conventionally, OCT is based on the interference of all photons, regardless ballistic or multiple scattering ones, from the sample beam with those from the reference beam.  In this study, we thus contruct the OCT signals directly the ballistic  (Fig.\ref{fig:figmonte}(b)) and multiple scattering  (Fig.\ref{fig:figmonte}(c)) photons, respectively.  We selected the center A-line (see the lateral position: $x=0.0mm$  in Fig.\ref{fig:figmonte}(a),(b),(c)), normalized by the top pixel,  to study the different characters of ballistic and multiple scattering signals , as shown in Fig.2d. Within a specific layer, both ballistic (red) and multiple (blue) OCT signals decrease linearly with the depth, however, ballistic OCT signals owns stronger intensity than multiple-scattering OCT signal, and there is more remarkable “bump” effect around the boundaries in ballistic OCT signals (Fig.2(d)).  It was noted that within the strong scattering RPE layer,  the ballistic signals decrease more remarkably with the depth than the multiple scattering signals.  The distinct log-linear trend and $SNR$ of ballistic and multiple scattering OCT signals along the A-line justify the separation of the two signals to  reveal more physiological details.

Fig.\ref{fig:figmonte}(e) shows the reflectance ratios of both ballistic and multiple scatterings to the hybrid signal along the same A-line. The ballistic OCT signal dominates in the superficial layers (retina) and gradually decrease with the depth, which becomes comparable with that of the multiple OCT signals in RPE layer, and then lost this dominance  beneath this layer.  Such a flip of the ballistic and multiple scattering signal ratios along this A-line thus help to interpret the RM-based separation could resolve the scatterings from different depth of the tissues.

We further look into the difference in eigenvalue ($s_i$) spectrum for the single scatterings, multiple scatterings and hybrid scatterings, respectively. Among the first seven largest eigenvalues ($s_i$) as shown in Fig.\ref{fig:figmonte}(f), the largest two for the single scattered photons are more than $300\sim500$ times higher than that for multiple scattered signals that is close to 0, however, such a difference quickly disappears starting from the 3rd largest eigenvalue.  Therefore, this pattern of eigenvalue spectrum in Fig.\ref{fig:figmonte}(f) rationalize the use of the first one or two largest eigenvalues to reconstruct the ballistic OCT signals mainly reflecting the superficial layer of the tissue, which provides the theory for our RMT based separation of OCT signals.

Fig.\ref{fig:figmonte}(d) also shows the reconstructed  OCT signals (green) based on RMT  only with the largest eigenvalue, which is pretty close to single scatterings (blue) in the superficial layers (<0.02cm). The difference signals (purple) between the original hybrid (not shown) and RMT reconstructed signal (green) is very close to the multiple scattering OCT signals (red).  MC simulation also shows that the single scattering signal (blue) is much lower than that of the multiple scattering due to more diffusion in deeper tissue (>0.02cm). This simulation thus also implies the usefulness of RMT reconstruction to improve the SNR if combined with the current processing methods for OCT signals.

\subsection{Improved SNR in in-vivo vis-OCT Data}

\begin{figure}
\centering\includegraphics[width=12cm]{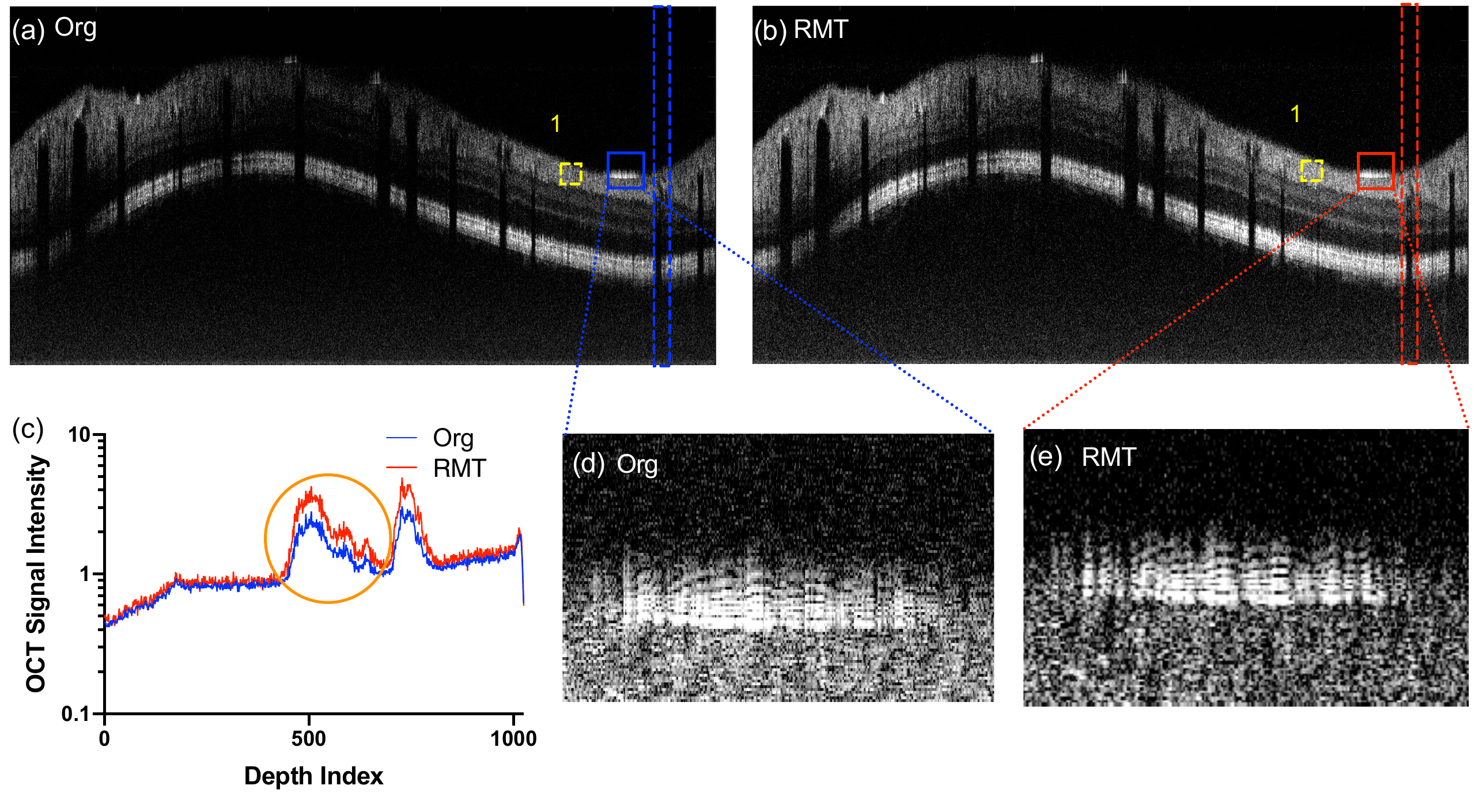}
\caption{Human retina circular scanning B-scan; (a) Original circular scanned OCT B-scan, (b) RMT reconstructed circular scanned OCT B-scan, (c). Compounding A-lines’ amplitude of (a) and (b)’s dashed ROIs, (d) Detail image of Original OCT on nerve head, (e) Detail image of RMT reconstructed OCT on nerve head}
\label{fig:fig129}
\end{figure}

\begin{figure}
\centering\includegraphics[width=12cm]{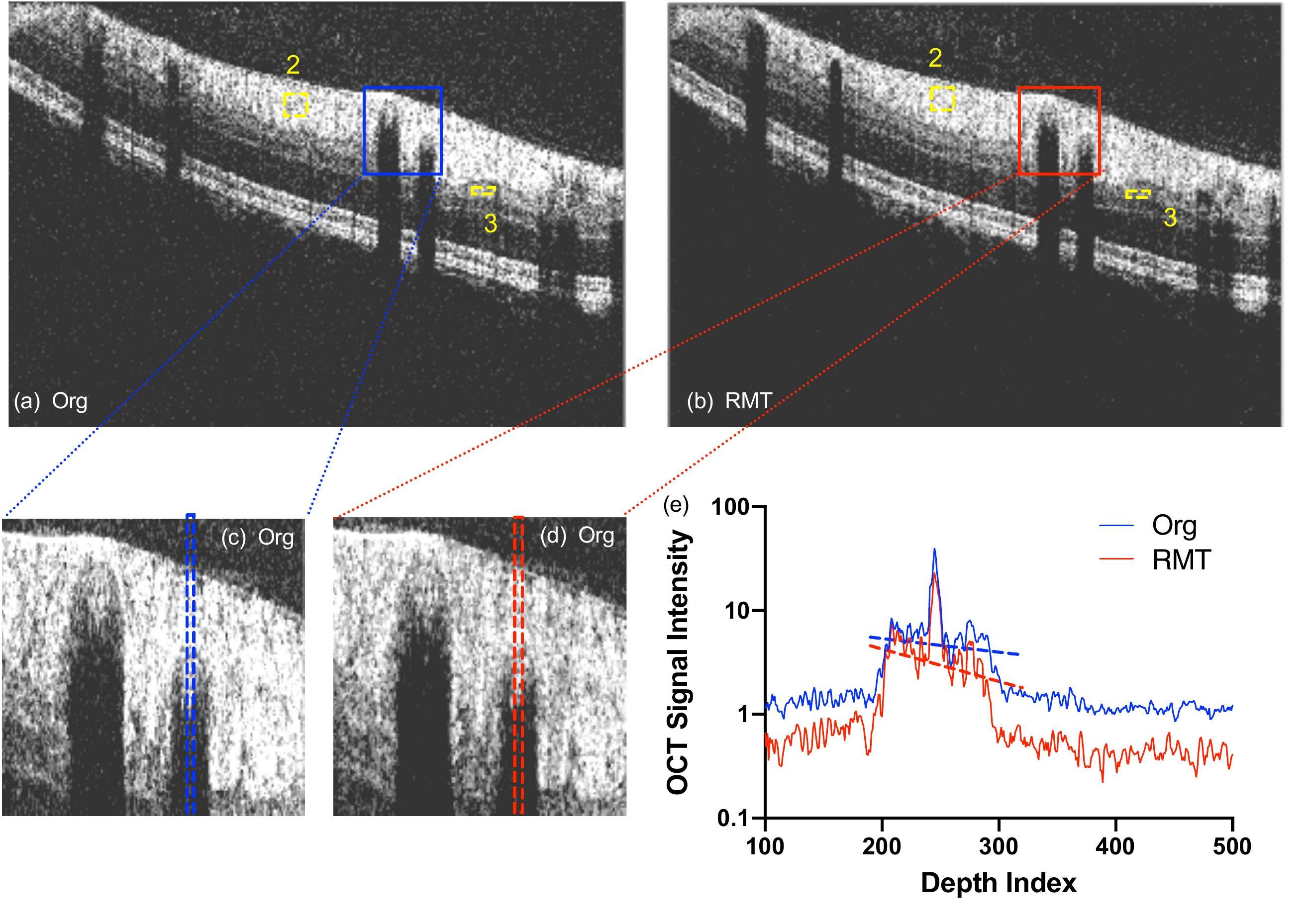}
\caption{Human retina raster scanning B-scan. (a) Original raster scanned OCT signal, (b) RMT reconstructed raster scanned OCT signal, (c) Zoom-in image including a vessel in (a), (d)RMT reconstructed (c), (e) Averaged A-lines’ amplitudes corresponding to the dashed A-lines’ bands in (b,c).}
\label{fig:fig000}
\end{figure}

To evaluate the denoising performance of the W-RMT, we analyze the data from an in-vivo vis-OCT experiment of human fundus. The experimental procedures have been described in Method session. The data include two types of OCT images from circular scanning (Fig.\ref{fig:fig129}(a)) and raster scanning (Fig.\ref{fig:fig000}(a)).

Two A-line bands are selected from Fig.\ref{fig:fig129}(a) and a window in Fig.\ref{fig:fig000}(a), respectively, to compare the SNR of the standard OCT signals (blue) and the reconstructed OCT signals (red) from the largest eigenvalue  of RM. The first peak in Fig.\ref{fig:fig129}(c)  is due to the reflection by the retinal surface, and the second peak corresponds to the strong scattering of the RPE. Shown in the circled region(orange), the RMT-based reconstruction reveals more distinguishable layers. Another highlight (color) is that RMT reconstruction still show significantly improved SNR in deeper layers. The RMT reconstructed image (Fig.\ref{fig:fig129}(e)) show more fine structure of retina than the orginal OCT signals (Fig.\ref{fig:fig129}(d)) with clear texture of the nerve head. 

In Fig.\ref{fig:fig000}(c). The RMT reconstructed OCT signal show near twice attenuation of the original OCT signal in Fig.\ref{fig:fig000}(e),  showing the suppressed multiple scattering in blood by discarding the majority of the small eigenvalues of RM.  We can further quantify the PSNR (Eq.11) and CNR (Eq.10) within the selected bands of A-lines.  In both circular and raster scanning, RM-based reconstruction significantly improve the PSNR with 4.19dB and 12.30dB, respectively. Table II shows that some regions(yellow boxes in Fig.\ref{fig:fig129} and Fig.\ref{fig:fig000}) have obvious improvement on details structure according to the CNR calculation.

\begin{table}
\centering
\caption{CNR(dB)}
\begin{tabular}{p{2cm}p{3cm}p{3cm}}
\arrayrulecolor{black}
\hline
Region.& Original & RMT-reconstructed \\&(Fig.3a, Fig.4a)&(Fig.3b, Fig.4b)   \\
\hline 
1(Fig.3)      & 0.9327	    &1.4940\\
2(Fig.4)      & 2.7595	    &4.2633\\
3(Fig.4)      & 0.9789	    &2.3604\\
\hline
\end{tabular}
\end{table}

\section{Discussions}
Our \emph{in vivo} experiment shows that RMT effectively separate the single scatterings from the raw vis-OCT signals, revealing more layered structure of the retina.

One of  the assumptions for RMT reconstribution is that the OCT signal is approximately \emph{i.i.d}, the estimation of its distribution could be affected by the choices of the number of A-lines as well as the scanning step. Larger size of RM always results in less estimations errors. In practice, the tissue inhomogeneity limit the number of A-lines used in calculation. In this study, we select 7 A-lines in the RM construction by checking the inhomogeneities in the acquired A-line signals. On the other hand, it should also be noted that high lateral resolution, or small scanning step, would introduce more correlation between the neighboring A-lines, which thus also violates the assumption of the independent sampling and lead to errors in RM. The scanning step of our OCT system , specifically designed for human retina imaging, was fixed at the A-line acquisition rate of 70 kHz. When applied to other tissues, the scanning step need be further  optimized, or we can  use an alternative strategy  of repeated scanning along the same location, which has been adopted in vis-OCT for $sO_2$ estimation[3].

Our RMT-based reconstruction considers the scattering effects only . Absorption, however, also plays an important role  in biological tissues, especially within the visible light spectrum.  The backscattered spectrum can be used to measure the tissue composition concentrations, e.g., oxygen and de-oxygen hemoglobin, according to the Beer-Lambert law. This conventional method  is based the assumption of single scatterings, which thus can be biased by the multiple scatterings in the signals, as used in vis-OCT[3]. Beer-Lambert's law has been directly applied in vis-OCT to estimate the $sO_2$ in large vessel which assuming in a single scattering regime. For tissue area and small vascular, there is no good method for $sO_2$ estimation currently due to the insufficient sensitivity. For the multiple scattering trajectories, the absorption procedure has higher probability to eliminate some of them and thus provides higher sensitivity. Further studies are needed to correctly extract the SO2 values in tissue areas.

We recognize that our isolation method proposed in this study works is in the scattering regime. For biological tissues, the absorption also plays an important role, especially for the light in visible spectrum. The absorption always eliminates trajectories which is more evident in multiple scattering paths. It enhances the ratio of single scattering component to the overall backscattered light. Meanwhile, the backscattered spectrum also provides the absorption spectrum which can be utilized to measure the tissue composition concentrations based on Beer-Lambert law. For example, traditional vis-OCT provides the estimation of oxygen and de-oxygen hemoglobin in blood vessel area. After isolation of single scattering components, the left signal contains more contributions from the multiple scattering components which will provides more accurate estimation for multiple spectral analysis to decode the tissue composition concentrations. At the same time, other robust method should be investigated to extract the multiple scattering component directly from the original backscattered light in vis-OCT.

\begin{figure}
\centering\includegraphics[width=12cm]{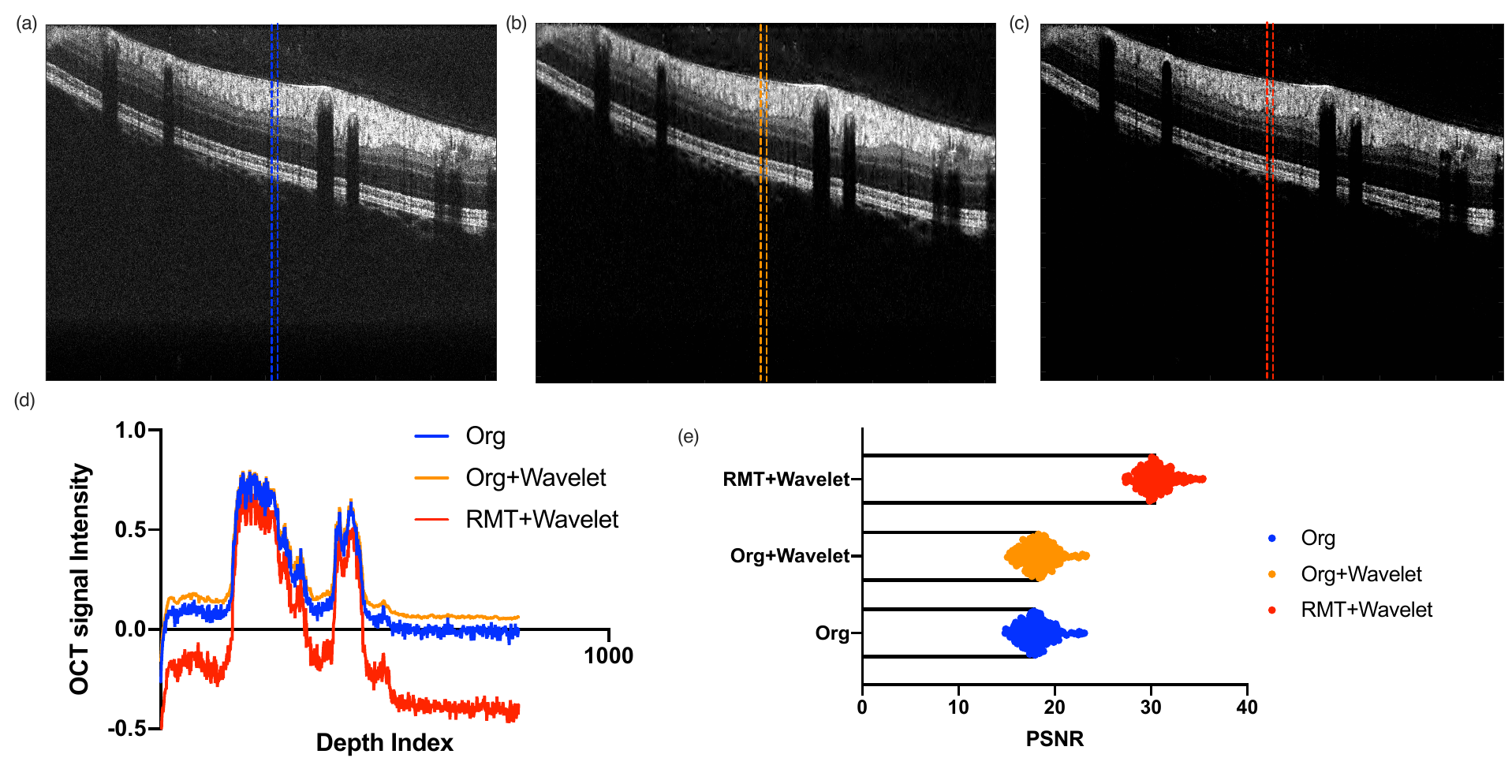}
\caption{(a)original image of human retina vis-OCT scanning B-scan; (b) denoising by using wavelet-based method on the orignal OCT image;(c)Denoising by using wavelet-based method on the RMT reconstructed OCT image .(d)Averaged OCT A-lines intensity within the selected band; (e)PSNR comparison betwen two denosing methods.}
\label{fig:fig5}
\end{figure}

It should be noted that RMT reconstruction was not proposed an denoising method to substite the current methods like wavelet-based denoising\cite{10.1117/12.734286,10.1117/1.3081543}, anisotropy curvelet transform\cite{Jian:10}, and sparsity-based denoising\cite{Fang:12}, but as a pre-processing for these existing denoise methods.  Fig.\ref{fig:fig5} shows how much the SNR can be improved by combining RMT and commonly used wavelet denoising method in OCT imaging.

Although the signals in this study are from vis-OCT which is a special case of SD-OCT, our method can also be applied to TD-OCT(Time Domain OCT) or SS-OCT(Swept Source OCT). Compared with SD-OCT,  TD-OCT filters the multiple scattering components more efficiently by scanning the reference arm for a better gating control.   Thus applying RMT to TD-OCT may not improve the SNR as good as that to SD-OCT. SS-OCT, however, sweeps the wavelength in time and  provides more independent coherent illuminations, which thus reduces the cross-talk between neighboring wavelengths and makes it more suitable for RMT reconstruction.  Most traditional OCT imaging techniques rely on the single scattering components. RMT provides a tool for seprating the single- and multiple scatterings, which facilitates more accurate structual and functional imaging of the tissues.  Besides, Doppler and polarization information can also be extracted more effectively from the RMT-reconstructed signals.

\section{Conclusion}
In conclusion, we proposed a statistical separation method based on W-RMT, through which the low-rank signal for single scatterings can be used to reconstruct vis-OCT signals with improved imaging quality.  MC  simulation confirmed the validity of the RMT-based separation of the single scattering and multiple scattering from the hybrid scatterings, which justifies the reconstructing high SNR OCT signals from the single scatterings only.  In an in-vivo experiment, RMT-based reconstruction could significantly improve the PSNR by 4.19dB and 12.30dB for the circular and raster B-scan signals,  respectively, with a significant value  for clinical diagnosis.  
The framework of the RMT-based reconstruction essentially enables studies on the scattering and/or the absorption difference in tissues with multiple layers. 
  
\begin{backmatter}

\bmsection{Funding}
This study is supported by the Shanghai Science and Technology Commission of Shanghai Municipality (Grant No. 19DZ2280300), Med-X Research Fund of Shanghai Jiao Tong University (YG2021QN16); and National Natural Science Foundation of China (NSFC No. 61876108).

\bmsection{Acknowledgments}

\bmsection{Disclosures}

\bmsection{Data availability} 
Data underlying the results presented in this paper are not publicly available at this time but may be obtained from the authors upon reasonable request.

\bmsection{Supplemental document}
No Supplementary materials. 

\end{backmatter}

\bibliography{ref.bbl}
\end{document}